First principles investigations of electronic and magnetic properties of new *rh.*-$B_{12}N_3$.

Samir F. Matar[†]

Lebanese German University, Sahel-Alma, Jounieh, Lebanon.

[†]Formerly at the University of Bordeaux, ICMCB-CNRS

Email: s.matar@lgu.edu.lb & abouliess@gmail.com

## Abstract

*Considering rhombohedral α-boron, rh-$B_{12}$ as a matrix hosting interstitials particularly triatomic linear ones as in $B_{12}C_3$ better known as $B_4C$, the subnitride $B_{12}N_3$ is proposed herein. The N3 triatomic linear alignment labeled N2-N1-N2 resembles that found in the ionic sodium azide $NaN_3$ characterized by a very short d(N1-N2)= 1.16 Å. Within DFT-based calculations $B_{12}N_3$ is found more cohesive than pristine $B_{12}$ with larger inter-nitrogen separation of d(N1-N2)= 1.38 Å. The N1-N2 elongation is explained from the bonding of the two terminal N2 with one of the two $B_{12}$ boron substructures forming "3B…N2-N1-N2…3B "-like complex. A resulting non-bonding charge density localized on central N1 leads to the onset of a magnetic moment of 1 $\mu_B$ in a half-metal ferromagnetic ground state illustrated by projections of the magnetic charge density and the site and spin electronic density of states DOS.*

**Introduction**

Opposite to many complicated structural forms of boron such as tetragonal β-$B_{192}$ [1, 2], rhombohedral α-$B_{12}$ (space group $R\bar{3}m$, N°166) is a relatively simple one. Depicted in Fig. 1a), α-$B_{12}$ presents two distinct crystal sites each with 6-fold multiplicity, i.e. B1 and B2 [3] (cf. Table 2) with the remarkable feature of an empty central void surrounded by B1 highlighted in Fig. 1b). The vacant space can be occupied by one, two and/or three interstitial atoms leading to a large family of compounds (cf. [4] for a review on boron with enumerated families).

Regarding triatomic interstitials, they are aligned and belong to two crystal sites (cf. Table 2c). They can be of the same chemical nature as in $B_{12}C_3$ also written $B_{12}${C-C-C}, known more commonly as $B_4C$ used for several applications pertaining to its large hardness [5] as well as comprising atoms of different chemical natures as in $B_{13}N_2$ prepared using high pressures [6] also written as $B_{12}${N-B-N} and recently studied $B_{12}${N-C-N} [4]. Noteworthy to mention here the occurrence of such hetero- and homo-triatomic alignments in much simpler rhombohedral structures of several ionic azides, such as $Na^{+1}N_3^{-1}$ or Na{N-N-N} [7] - cf. Fig. 1c), and calcium cyanamide $Ca^{+2}(CN_2)^{-2}$ also expressed as Ca{N-C-N} [8]. We also note the occurrence of linear cyanamide anion in hexagonal space group $P6_3mmc$ N°194 such as Fe{N-C-N} [9].

Besides well known $B_4C$, the $B_4N$ composition has gained some interest in recent years mainly as a potential abrasive [10]. A cubic form based on antiperovskite-like γ'-$Fe_4N$ was theoretically proposed [11] by replacing Fe by B, while keeping nitrogen at the cube center. Also in the search of novel ultra-hard boron nitrides under pressure, such composition was proposed by a Chinese group, $B_{24}N_6$ in a large cubic cell from structure prediction calculations [12]. The characteristic feature of the structure is the favoring of B-N bonds with a short distance of 1.57 Å while d(B-B)=1.80 Å. The cubic cell has no close N-N neighbors as in $NaN_3$ azide and hence no similarity with presently studied compounds.

In view of the above introduced context of $B_{12}$ as a host matrix of interstitials on one hand and of the aligned triatomic N-N-N in known azide compounds on the other hand, a stable form of rhombohedral $B_4N$ is proposed herein in the formulation $B_{12}N_3$ (Fig. 1d) with a study of its electronic structure. In Fig. 1c) and 1d) the azide entity is labeled N2-N1-N2 because N1 and N2 belong to two Wyckoff sites as explicated in Table 1b) and 1c).

Specifically it will be shown that $B_{12}N_3$ is cohesive, and possesses particular electronic and magnetic properties due to the interaction of N2 with one of the two boron substructures B1, by forming a complex-like entity "3B1···N2–N1–N2···3B1" separated from the other boron substructure (B2) and leading to a peculiar magnetic polarization on nitrogen N1.

The framework of the investigation is the quantum density functional theory DFT [13, 14] which was used to examine linear cyanamide {N-C-N} inserted in $B_{12}$ [4].

1. **Computational framework**

The search for the ground state structure and energy was carried out using calculations based on the DFT within the plane-wave VASP code [15, 16] and the projector augmented wave (PAW) method [16, 17] for the atomic potentials accounting for all valence states especially in regard of the light elements B and N. The DFT exchange-correlation effects were considered with the generalized gradient approximation (GGA) [18]. The conjugate-gradient algorithm [19] was used in this computational scheme to relax the atoms onto the ground state. The tetrahedron method with Blöchl et al. corrections [20] and Methfessel-Paxton [21] scheme was applied for both geometry relaxation and total energy calculations. Brillouin-zone (BZ) integrals were approximated using a special **k**-point sampling of Monkhorst and Pack [22]. The optimization of the structural parameters was performed until the forces on the atoms were below 0.02 eV/Å and all stress components less than 0.003 eV/Å$^3$. The calculations were converged at an energy cut-off of 500 eV for the plane-wave basis set concerning the **k**-point integration with a starting mesh of 6×6×6 up to 12×12×12 for best convergence and relaxation to zero strains.

Properties related with the electron localization were obtained from real-space analysis of electron localization function (ELF) according to Becke and Edgecomb [23] as initially devised for Hartree–Fock calculations then adapted to DFT methods as based on the kinetic energy in which the Pauli Exclusion Principle is included by Savin et al. [24] ELF = $(1+ \chi_\sigma^2)^{-1}$ with 0 ≤ ELF ≤1, i.e. ELF is a normalized function. In this expression the ratio $\chi_\sigma = D_\sigma/D_\sigma^0$, where $D_\sigma = \tau_\sigma - ¼(\nabla\rho_\sigma)^2/\rho_\sigma$ and $D_\sigma^0 = 3/5\ (6\pi^2)^{2/3}\rho_\sigma^{5/3}$ correspond respectively to a measure of Pauli repulsion ($D_\sigma$) of the actual system and to the free electron gas repulsion ($D_\sigma^0$), and $\tau_\sigma$ is the kinetic energy density. In the post-treatment process of the ground state electronic structures, the total charge density "CHGCAR", as well as the magnetic charge density "CHGCAR_magn" are illustrated.

From the geometry of the ground state structures in NSP and SP configurations, the electronic site and spin projected density of states (PDOS) were obtained using DFT-based full potential augmented spherical wave (ASW) method [25,26] and the GGA for the XC effects [17]. In the minimal ASW basis set, the outermost shells were chosen to represent the valence states and the matrix elements. They were constructed using partial waves up to $l_{max}$ + 1 = 2 for B and N. Self-consistency was achieved when charge transfers and energy changes between two successive cycles were respectively $\Delta Q < 10^{-8}$ and $\Delta E < 10^{-6}$ eV. The BZ integrations were performed using Blöchl et al. [20].

## 2. Calculations and discussion of the results

### a- Trends of cohesive energies

A preliminary step was to examine the above cited compounds from their respective cohesive energies obtained considering total spin configuration NSP within unconstrained, parameter-free, successive self-consistent sets of calculations at an increasing number of **k**-points. The cohesive energies deducted from subtracting the constituents' atomic energies are averaged as per atom to enable comparison between the different stoichiometries.

Cubic $B_4N$ based on γ'-$Fe_4N$ structure [11] was calculated with a positive cohesive energy, i.e. non cohesive, while $B_{24}N_6$ [12] was calculated cohesive with $E_{coh.}$/at. = -1.25 eV. Pure $rh.$-$B_{12}$ has $E_{coh.}$/at. = -1.15 eV and $E_{coh.}$/at.($rh.B_{12}C_3$)= -1.51 eV showing better cohesion within $B_{12}$ due to the embedding of $C_3$ whereby B-C bonds add up to the stability versus pure hosting $B_{12}$. This also stands for presently considered dodecaboron azide with $E_{coh.}$/at.($rh.B_{12}N_3$)= -1.19 eV. The B1-C bonding in $B_4C$ is hence more favorable than B1-N one. However its cohesive energy comes close to $B_{24}N_6$ which mainly comprises B-N shortest bonds and no linear $N_3$. Lastly, ionic $NaN_3$ was found largely cohesive with $E_{coh.}$/at.($rh.NaN_3$) = -2.51 eV.

b- Calculations and results.

- Geometry optimization.

Focusing on $B_{12}$, $NaN_3$, and $B_{12}N_3$ total spin, non spin polarized NSP calculations were carried out for unconstrained geometry optimizations to relax the atoms onto the ground state. The structural results are given in Table 1.

For $B_{12}$ (Table 1a), a relatively good agreement between experimental and calculated values is obtained for the lattice constants and the atomic positions for B1 and B2 substructures. $NaN_3$ is calculated with d(N1-N2) slightly larger than the very short experimental distance of 1.16 Å which leads subsequently to smaller d(Na-N2) of 2.40 Å. Nevertheless the major outcome sought here is to confront the changes in the linear $N_3$ upon its embedding in $B_{12}$ host. With respect to $B_{12}$, the additional one N1 and two N2 are the same as in $NaN_3$, i.e. N1(1*b*) at ½, ½, ½ and N2(2*c*) at *x, x, x* with *x* = 0.442. As a result, a slightly smaller starting magnitude of *x* ~0.400 was chosen *ad hoc* because in such less ionic environment d(N-N) is expected to be larger than 1.16 Å characterizing $NaN_3$. Note that such a choice does by no means bias the calculations which are self consistent and parameter free. As a matter of fact the calculated *x*(N2) amounts to 0.388. Hence, a key information is obtained regarding the change of N1-N2 separation: d(N1-N2) = 1.38 Å, meaning a significant extension versus the value in $NaN_3$. The other short distance is d(B1-N2) =1.57 Å and B1-B1 separation is enlarged with respect to B2-B2 other substructure. Then the N1-N2 elongation results from the formation of complex-like "3B1…. N2–N1–N2…3B1" where each of the two N2 atoms bind with 3 B1 leading to less interaction with N1 and to the change of the nature of the N1-N2 bond by departing from the azide triple bond-like (as observed in the molecular state), to a double-bond like one involving the creation of non bonded electrons on N1. The non bonding charge density together with the separation of N1 from N2 leading to isolated non bonding charge density will be responsible for the trend to develop magnetization which requires localization of the electronic states as shown below. From these changes the $B_{12}N_3$ structure shown in Fig. 1d) is significantly different from $B_{12}$ one in Fig. 1a). Another feature is in the significant increase of $\alpha_{rh.}$ =63.50° with respect to its value in pristine $B_{12}$, i.e. $\alpha$ = 58° due to sterical effect relevant to the 'opening' of the $B_{12}$ rhombohedron to welcome the linear $N_3$ in the void depicted in Fig. 1b). This also stands for $B_{12}C_3$ ($B_4C$) with an even larger opening: $\alpha_{rh.}$ =65.5° [5]. Clearly interstitials greatly modify $B_{12}$ skeleton and are likely to bring new physical properties.

- Total and magnetic charge density.

The charge density resulting from the self-consistent calculations for $NaN_3$ and $B_{12}N_3$ are illustrated in Fig. 2a) and 2b) respectively with 3D grey envelopes and 2D slices. Blue and green color zones correspond to zero and free electron-like spots. In $NaN_3$ the linear $N_3$ shows triple spherical shape of grey envelopes, it is embedded in a green area of free electron-like behavior surrounding it, but an insulating character can be expected from the dominant blue zone separating $N_3^{-1}$ from $Na^{+1}$.

Oppositely, in $B_{12}N_3$ there is a clear skewing of the two terminal $N_2$ charge density towards the three neighboring B1 en each side of linear N2-N1-N2, exhibited by a flattening of the grey envelopes. Regarding the 2D slice, there can be observed a continuous free electron like green area extending to all the constituents including B2 pointing to the bonding. A metallic character can be expected. Numerically, this is supported by the odd number of valence electron count (VEC) of 51 electrons, i.e., by accounting for B $(2s^2,2p^1)$ ➔ VEC = 3 and $N(2s^2,2p^3)$ ➔ VEC = 5, VEC($B_{12}N_3$)= 12×3 + 3×5 = 51 while an even VEC count characterizes $B_{12}$ with VEC($B_{12}$)= 36, i.e. 12×3 = 36. Also with VEC($B_{12}C_3$) = VEC(B12) + 3×4 = 48, one expects closed shells for $B_{12}$ and $B_{12}C_3$. Besides the expectation of a paramagnetic behavior for $B_{12}N_3$, and in so far that DFT is implicitly a zero temperature (0K) theory with no account for entropy, further calculations assuming spin polarization SP could lead to a possible onset of magnetic polarization in the ground state.

Indeed starting from the non spin polarized calculations and accounting for two spin channels, i.e. spin up ↑ and spin down ↓ the calculation energy results give a slight difference between the two magnetic configurations favoring the SP one with ΔE(SP-NSP) = -0.51 eV, thus providing a magnetic ground state and a magnetic moment $M_{Tot.}$ = 1 $\mu_B$ (Bohr Magneton) assigned to the whole cell (cf. section c- SP DOS). The lattice constants and parameters differ only slightly from NSP ones.

Since the magnetic moment is calculated from the difference between ↑(populations) and ↓(populations) corresponding to two charge densities, the difference between them provides the magnetic charge density localization. This is graphically shown in Fig. 2c). The 2D slice crossing the same plane as in Fig. 2b) shows mainly a blue area of no magnetization except around central N1 with a red central area with a torus-like 3D grey volume around it and an extension to a large green area whose traces are also found in the neighborhood of N2, corresponding to residual magnetic charges. Then to a large extent the magnetic moment is mainly carried by N1 with residues on neighboring N2.

This is further explained upon plotting the site and spin projected density of states DOS in the next sections.

- Electron localization

Further illustration regarding the localization of the electrons and the corresponding bonding is obtained from the 2D and 3D plots of the ELF in Fig. 3. As introduced in the computation section, ELF is a normalized function with $0 \leq ELF \leq 1$. The ruler indicates the color code with 0 corresponding to zero localization with blue zones, ELF= ½ for free-electron-like behavior with green zones and ELF = 1 for full localization with red zones. Afore mentioned "3B1…. N2–N1–N2…3B1" complex entity can be seen at the center of the rhombohedron with grey volumes of localization found between B1 and N as well as a torus of non-bonded electron localization around N1 thus showing that the magnetic charge density plot in Fig. 2c) correspond to non bonded electrons as being at the origin of the magnetization. In spite of its large separation from B1, the B2 substructure keeps connected with the B1 substructure with green ELF zones of free electron like behavior extending throughout the structure, letting suggest a rather conducting electronic system which will be confirmed from the DOS.

c- Site and spin projected density of states

The specific role of each chemical constituent in $B_{12}N_3$ can be assigned based on the projection of the electronic density of states DOS in the two magnetic configurations, NSP and SP respectively obtained within the full potential augmented spherical wave (ASW) method [25,26] using the GGA gradient functional for the XC effects [17].

Fig. 4a) shows the plot of the NSP DOS. In this panel and in the SP following one, the energy zero along the *x*-axis is with respect to the Fermi level $E_F$.

There are three main energy regions: in the lowest energy part N2(2s) states are found split into two narrow PDOS; the lower energy peak is due to the s-like N1-N2 quantum mixing, while at -19 eV one finds the N2 mixing with less electronegative B1 s-like sates. The energy region {-15; -1 eV} is mainly dominant with p states showing B1-B2 quantum mixing translating the bonding as well as B1-N through the p states in the lower energy part where at -14 eV one finds N1-N2 similar PDOS. A most interesting feature appears at the Fermi level crossed by a large N1 PDOS versus much smaller intensity PDOS arising from N2 and B1. High intensity NSP DOS@$E_F$ signals instability of the electron system (of N1-2p) in such total spins configuration [27]. Lowering of the electronic system energy is expected upon accounting for two spin channels, i.e., within spin-polarized SP calculations which were carried out subsequently. Above $E_F$, at 2.5 eV, the empty conduction band CB mirrors the two features of N1-N2 and B1-B2 quantum mixings observed below $E_F$ with the VB.

Fig. 4b) shows the site and spin -projected SP-PDOS which mirrors to some extent the NSP panel, especially regarding the quantum mixing (bonding). The energy difference with respect to NSP is ΔE(SP-NSP) = -0.49 eV, a value close to the magnitude obtained from VASP calculations above, thus confirming the trend towards SP ground state.

The major difference versus NSP configuration is the energy downshift of majority spins (↑) and the energy up-shift for the minority spins (↓), mainly for the lower energy part around -20 eV involving N2 and B1 s-states and at $E_F$ where a ~2eV band gap is observed for (↑) spins while minority spins (↓) a large N1-PDOS peak is on top of $E_F$ signaling a half metallic magnet with M = 1 $\mu_B$. Note that the magnetic polarization involves mainly N1 from the large PDOS but also to a much lesser extent, the other states as N2 and partially B1 due to the quantum mixing.

Lastly additional ASW anti-ferromagnetic calculations starting from a doubled cell containing two magnetic subcells (UP and DOWN) led to an increase of the energy of the SP configuration, letting assume that $B_{12}N_3$ is a half-metal ferromagnet just like the property observed for the only ferromagnetic oxide $CrO_2$ where $Cr^{II}$ $3d^2$ [28, 29] shows a magnetization of 2 $\mu_B$.

### 3- Conclusions

Based on experimental observations of rh.-$B_{12}$ compounds such as well known $B_4C$, its expression as $B_{12}${C:C:C} lets identify triatomic linear entity shown for a series of compounds presented in a recent review [4]. By letting $B_{12}$ host triatomic linear azide {N:N:N} in the rhombohedral system, leads to propose $B_{12}N_3$ (3 $B_4N$) characterized as cohesive with a calculated magnetic ground state identified from energy differences. Magnetic polarization with M = 1$\mu_B$ develops on central N1 found to be less bonded to N2 in an elongated N2-N1-N2 with $d_{N1-N2}$ = 1.38 Å versus $d_{N1-N2}$ = 1.16 Å in the ionic $NaN_3$. The binding between each terminal N2 with B1, one of the two boron substructures, forms a "3B1…N2-N1-N2…3B1"-like complex, separated from the other boron substructure B2 as illustrated from total and magnetic charge density as well as from the electron localization ELF. A magnetic torus around relatively isolated N1 from N2 coincides with non bonded electron localization (ELF) around it.

**Conflict of interest**

The author declares non conflict of interest regarding this work and its results.

TABLES

**Table 1**. Experimental and calculated crystal data of the Rhombohedral compounds under consideration. Space group $R\bar{3}m$, N°166.

a) $B_{12}$ [3]

$a_{rh.}$= 5.057 (4.98)Å; $\alpha$ = 58.06° (58.47°).

| Atom | Wyckoff | x | y | z |
|---|---|---|---|---|
| $B_1$ | 6h | 0.221 (0.223) | x | -0.370 (-0.371) |
| $B_2$ | 6h | 0.010 (0.011) | x | 0.653 (0.650) |

d(B1-B1) = 1.65 Å; d(B2-B2) = 1.77 Å.

b) Experimental [8] and (calculated) rhombohedral $NaN_3$.

$a_{rh.}$= 5. 49 (5.47) Å; $\alpha$ = 38.93° (36.0°);

| Atom | Wyckoff | x | y | z |
|---|---|---|---|---|
| Na | 1a | 0. | 0……… | 0 |
| N1 | 1b | ½ | ½ | ½ |
| N2 | 2c | 0.442(0.422) | x | x |

d(Na-N2) = 2.51 (2.40) Å; d(N1-N2) = 1.16 (1.18 )Å.

c) $B_{12}N_3$ calculated in the ground state SP configuration.

$a_{rh}$ =5.187 Å; $\alpha$ =63.50°

| Atom | Wyckoff | x | y | z |
|---|---|---|---|---|
| $B_1$ | 6h | 0.804 | 0. 804 | 0.319 |
| $B_2$ | 6h | 0.002 | 0. 002 | 0.332 |
| N1 | 1b | ½ | ½ | ½ |
| N2 | 2c | 0.388 | x | x |

Volume (Å$^3$) = 107.07

Distances in Å.

d(N1-N2)=1.38;
d(N2-B1) =1.57;
d(B1-B1)= 2.64;
d(B2-B2) =1.80;
d(B1-B2) =1.72.

FIGURES

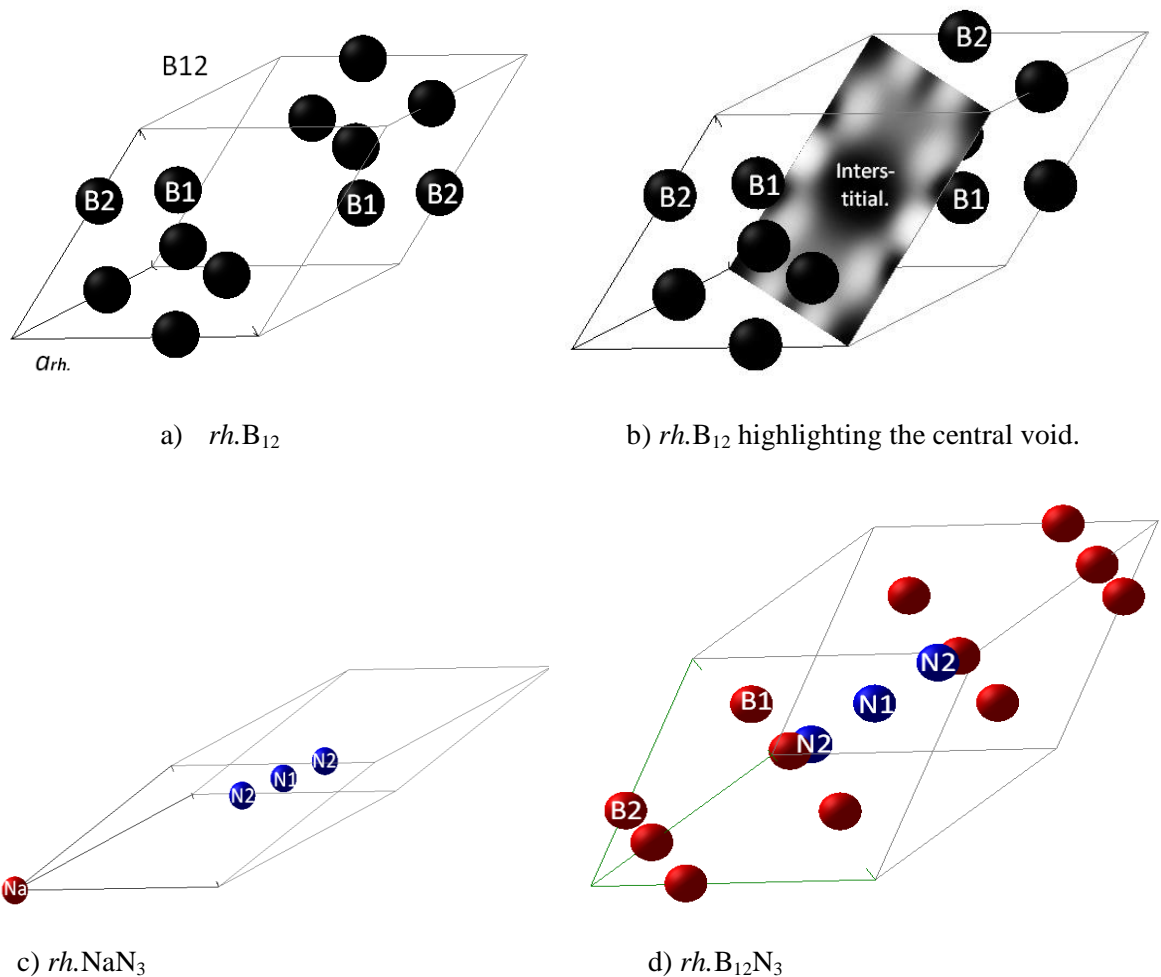

a)  *rh.*$B_{12}$

b) *rh.*$B_{12}$ highlighting the central void.

c) *rh.*$NaN_3$

d) *rh.*$B_{12}N_3$

Fig. 1 Sketches of the rhombohedral crystal structures. a) α-boron $B_{12}$; b) the interstitial void (large black spot) in $B_{12}$ highlighted with an electron localization slice (cf. ELF in text), c) $NaN_3$, and d) $B_{12}N_3$ with the interstitial linear N-N-N highlighted with a labeling N2-N1-N2 used also in sodium azide d).

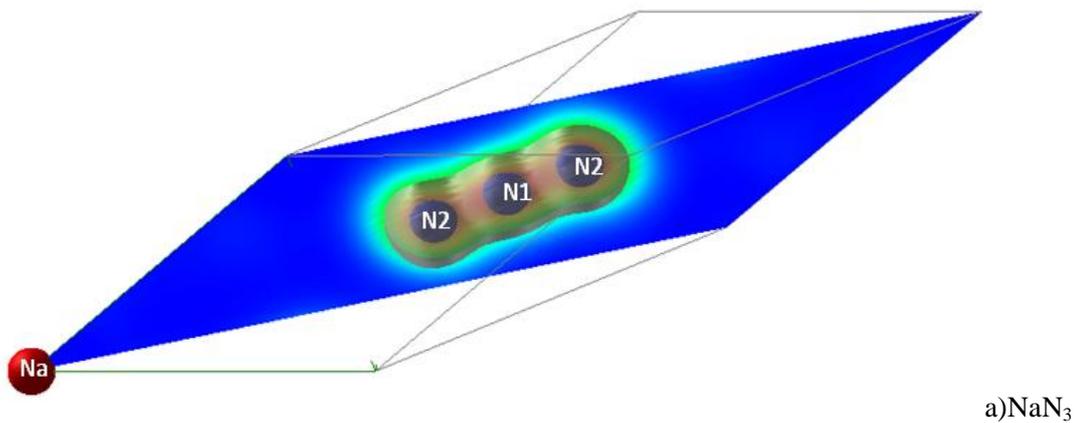

a) NaN$_3$

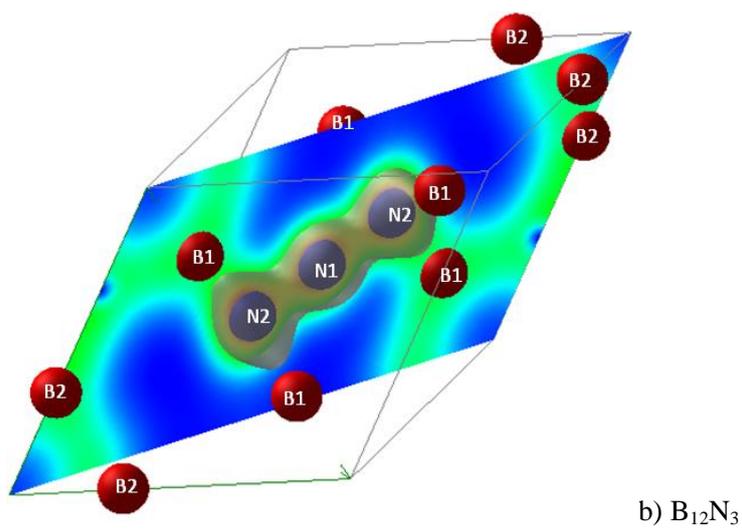

b) B$_{12}$N$_3$

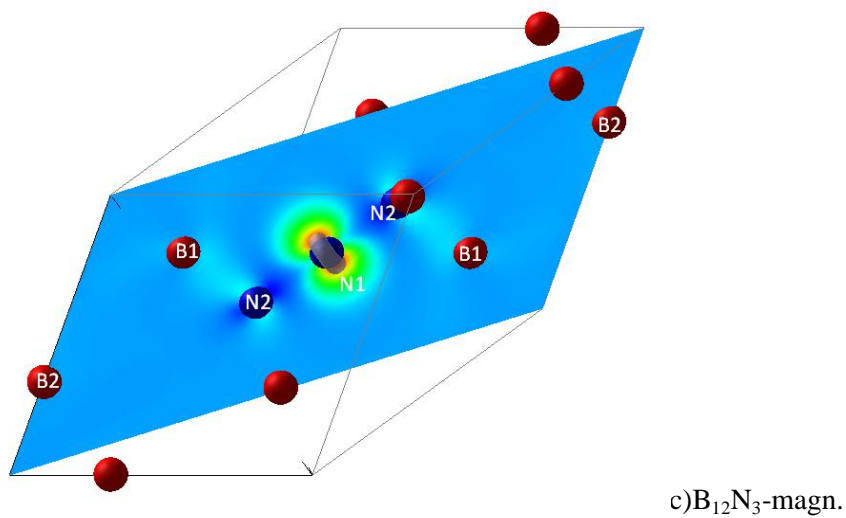

c) B$_{12}$N$_3$-magn.

Fig. 2. 3D grey envelopes of charge density and 2D slices. a) NaN$_3$ and b) B$_{12}$N$_3$.

c) shows the magnetic charge density in the form of a torus prevailing on the central nitrogen N1 in B$_{12}$N$_3$ and corresponding to non bonded electrons.

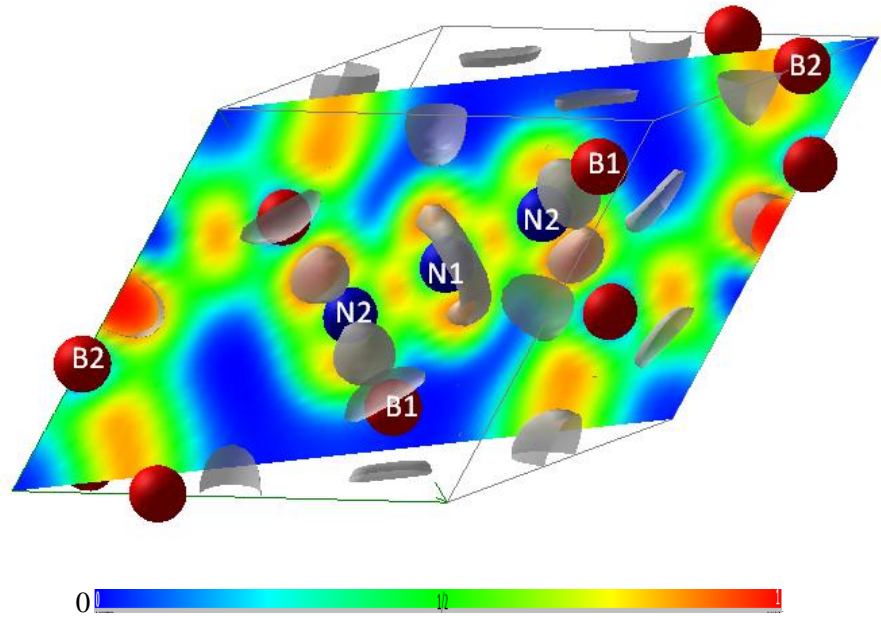

Fig. 3. Electron localization function ELF in $B_{12}N_3$ showing 3D envelopes (grey volumes). For 2D slice-plane the ruler indicates the color code from 0 (no localization) to 1 (full localization) with ELF=½ corresponding to free-electron-like localization.

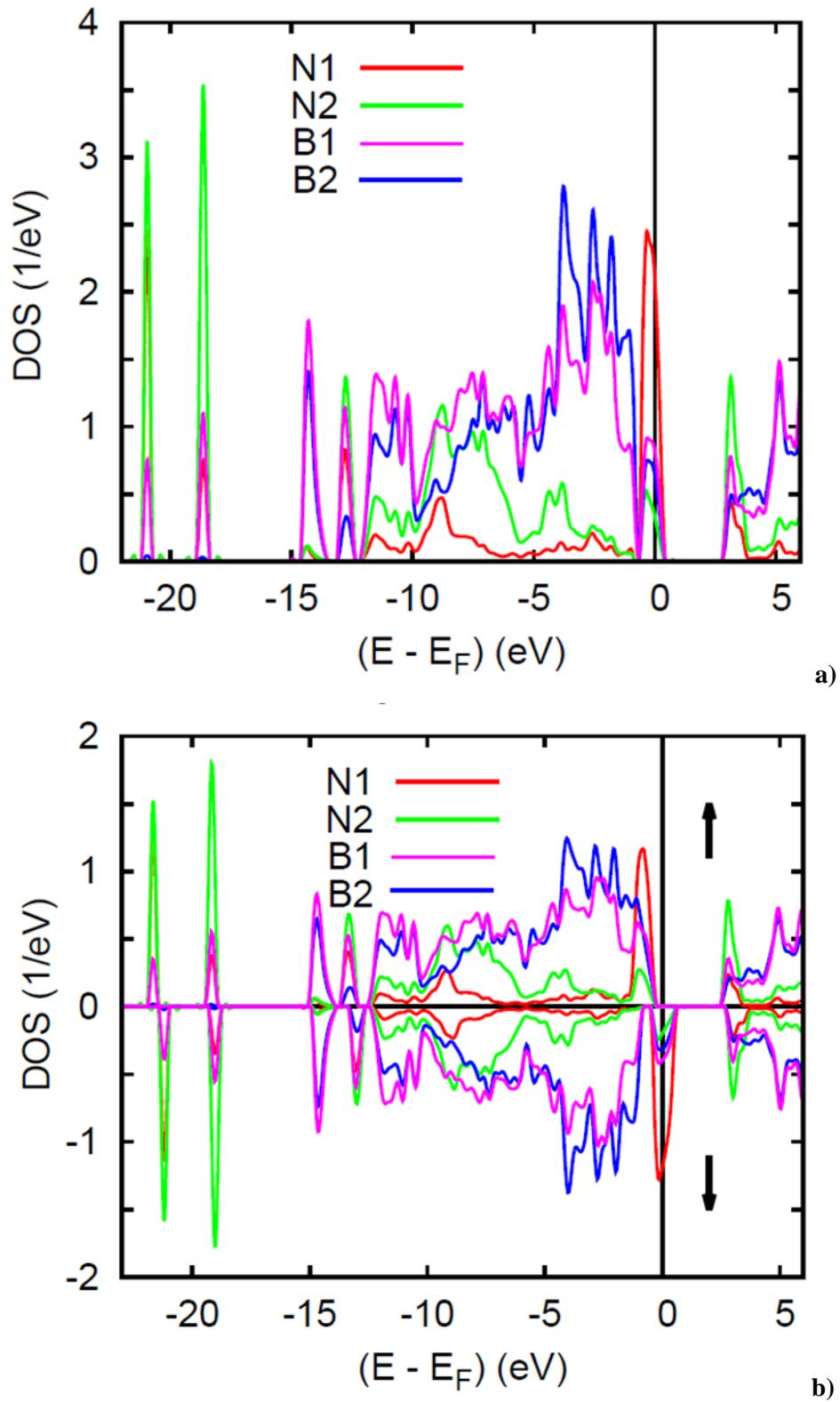

Fig. 4. $B_{12}N_3$: Electronic density of states DOS. a) NSP site projected and b) SP site and spin projected.